\documentclass[%
preprintnumbers,
amsmath,amssymb,
aps,
prb,
]{revtex4-1}
\usepackage{graphicx}
\usepackage{dcolumn}
\usepackage{bm}
\usepackage{hyperref} 
\hypersetup{colorlinks=true, citecolor=blue, urlcolor=blue, linkcolor=blue}

\usepackage{blindtext}
\usepackage[a4paper, total={6in, 8in}]{geometry}

\begin{document}

\title{When Plasmonic Colloids Meet Optical Vortices - A Brief Review}


\author{G.V. Pavan Kumar}
\email{pavan@iiserpune.ac.in}
\affiliation{
 Department of Physics, Indian Institute of Science Education and Research, Pune - 411008, India  
}

\begin{abstract}
Structured light has emerged as an important tool to interrogate and manipulate matter at micron and sub-micron scale. One form of structured light is an optical vortex beam. The helical wavefront of these vortices carry orbital angular momentum which can be transferred to a Brownian colloid. When the colloid is made of metallic nanostructures, such as silver and gold, resonant optical effects play a vital role, and the interaction leads to complex dynamics and assembly. This brief review aims to discuss some recent work on trapping plasmonic colloids with optical vortices and their lattices. The role of optical scattering and absorption has important implications on the underlying forces and torques, which is specifically enunciated. The effect of spin and orbital angular momentum in an optical vortex can lead to spin-orbit coupling dynamics, and these effects are highlighted with examples from the literature. In addition to assembly and dynamics, enhanced Brownian motion of plasmonic colloids under the influence of a vortex-lattice is discussed. The pedagogical aspects to understand the interaction between optical vortex and plasmonic colloids is emphasized.
\end{abstract}
\maketitle

\tableofcontents

\section{Introduction}
Vortices can be found in a variety of natural situations such as flowing water and spinning galaxies, and in our everyday lives.
They span a wide range of physical phenomenon and have fascinated humanity for a long time \cite{lugtVortexFlowNature1983}.
A laser beam too can propagate as an optical vortex.
As with all vortices, optical counterparts too carry angular momentum.
Interestingly, optical vortices can also trap and tweeze micro and nanoscale objects such as colloids made of various materials. One such category is plasmonic nano-colloids. 
These Brownian objects are sub-wavelength, metallic nanoparticles dispersed in a fluid. Due to the localized surface plasmon-polaritons, which are essentially electronic charge density oscillations driven by light, these colloids strongly absorb and scatter light at visible wavelengths. Therefore, resonant optical effects come in to play. When such nano-colloids interact with optical vortex traps, they show interesting dynamic effects. This short review aims to highlight interaction of optical vortex traps with plasmonic nano-colloids, and the emergent assembly and dynamics.

\subsection{Plasmonic Colloids}
One of the important aspect of reducing the size on an object to nanoscale is that the surface to volume ratio is increased enormously \cite{schommersBasicPhysicsNanoscience2018, wolfNanophysicsNanotechnologyIntroduction2008,natelsonNanostructuresNanotechnology2015}.
This means, majority of the physical response, and hence physical properties from the nano-object is arising due to the atoms residing at the surface.
This surface sensitivity is what makes nanotechnology \cite{schommersBasicPhysicsNanoscience2018} so powerful because it is the interface that plays a vital role in interactions. 
Optical properties of nano-objects too are altered due to the reduction in size, as shown historically by Micheal Faraday \cite{faradayBakerianLectureExperimental1857,tyndallFaradayDiscoverer1894}. 
The case in point is metal nano-colloids \cite{natelsonNanostructuresNanotechnology2015} made of coinage metals such gold, silver, copper, platinum and palladium. 
Specifically, gold and silver colloids, in the size range of $10$ nm to $500$ nm, have strongly optical response \cite{kellyOpticalPropertiesMetal2003a, kreibigOpticalPropertiesMetal1995,bohrenAbsorptionScatteringLight2008,lakhtakiaNanometerStructuresTheory2004} when illuminated by visible light\cite{henryCorrelatedStructureOptical2011}.
The absorption and scattering of light by these colloids are predominantly influenced by a strong resonance of electronic charge density waves, commonly termed as surface plasmon polaritons (or in short, surface plasmons) \cite{maierPlasmonicsFundamentalsApplications2007, lalNanoopticsSensingWaveguiding2007}.
The plasmons can be localized or propagating in nature, and therefore optical excitation can be converted into localized or propagating source by tailoring the geometry of the object \cite{barnesSurfacePlasmonSubwavelength2003,poloElectromagneticSurfaceWaves2013,polojr.SurfaceElectromagneticWaves2011} or by fabricating them in certain unconventional geometries \cite{deaberasturiModernApplicationsPlasmonic2015} including crystals and quasi-crystals\cite{achantaPlasmonicQuasicrystals2015, achantaSurfaceWavesMetaldielectric2020, patraQuasiperiodicAirHole2016}.\
Importantly, gold colloids of size $100$ nm can facilitate strong optical scattering due localized surface plasmons \cite{henryCorrelatedStructureOptical2011} , whereas elongated structures such as silver nanowires of width $150$ nm and length of 5 $\mu m$ can facilitate propagating surface plasmons \cite{barnesSurfacePlasmonSubwavelength2003, lalNanoopticsSensingWaveguiding2007}. 
In addition to radiative processes, plasmonic colloids can also offer non-radiative interactions, and can be used as heat sources. 
This 'thermoplasmonic' effect \cite{baffouThermoplasmonicsHeatingMetal2017} of plasmonic colloids has turned out to be important in applications including photothermal therapy and optothermal trapping and manipulations \cite{baffouApplicationsChallengesThermoplasmonics2020, lalNanoshellEnabledPhotothermalCancer2008}. 
Thus, plasmonic colloids offer a large variety of optical and optothermal effects that can be harnessed not only in fundamental studies of light-matter interaction, but also in applications including nano- and bio-photonics. 
In this review, we mainly focus on plasmonic colloids that are dispersed in water, and are undergoing Brownian motion.
Due to their pronounced optical absorption and scattering, their interaction with optical vortex leads to interesting translation and rotation dynamics, which will be discussed in this review.  

\subsection{Optical Tweezers}
Experimental exploration of optical trapping was initiated by Ashkin in 1970s \cite{ashkinAccelerationTrappingParticles1970, ashkinHistoryOpticalTrapping2000}. 
The first report on using radiation pressure to accelerate and trap particles \cite{ashkinAccelerationTrappingParticles1970} was thanks to the emergence of lasers.
The radiation pressure of the laser beam was also show to levitate objects \cite{ashkinOpticalLevitationRadiation1971} which created interest among scientists who were looking for micromanipulation tools.
A definitive breakthrough happened in 1986 when Ashkin and coworkers reported how a single laser can facilitate gradient optical forces \cite{ashkinObservationSinglebeamGradient1986} that can "tweeze" a dielectric particle in a liquid.
This was the origin of optical tweezer, and has been one of the outstanding optical tools harnessed for various applications \cite{jonesOpticalTweezersPrinciples2015}.
Interestingly, about the same time, Chu, Ashkin and coworkers used their knowledge on radiation pressure to trap atoms \cite{chuExperimentalObservationOptically1986}, which further lead to laser cooling \cite{chuNobelLectureManipulation1998, phillipsNobelLectureLaser1998}.
Ashkin and coworkers further pushed the applicability of optical tweezer to trap and manipulate single biological cells \cite{ashkinOpticalTrappingManipulation1987}, bacteria and viruses \cite{ashkinOpticalTrappingManipulation1987a} and internal of cells\cite{ashkinInternalCellManipulation1989} using lasers at infrared wavelength, which introduced optical tweezers to biologists.
Systematic origin of forces in geometrical optical regime \cite{ashkinForcesSinglebeamGradient1992} was identified in the context of optical trap, which further strengthened the fundamentals of optical tweezers.
A good overview of historical account of optical trapping and tweezing can be found in reviews of Ashkin  \cite{ashkinHistoryOpticalTrapping2000, ashkinOpticalTrappingManipulation2006}.\

 Around 1994, Svoboda and Block \cite{svobodaOpticalTrappingMetallic1994} showed how metallic Rayleigh particles, which are essentially plasmonic colloids, can be trapped using laser beams. 
This was one of the definitive papers to show how laser beams can be harnessed to trap and manipulate resonant nano-objects and motivated research in optical trapping of nano-colloids \cite{dienerowitzOpticalManipulationNanoparticles2008, lehmuskeroLaserTrappingColloidal2015, dholakiaColloquiumGrippedLight2010, maragoOpticalTrappingManipulation2013}.
Optical tweezers have emerged as indispensable tools \cite{grierOpticalTweezersColloid1997, grierRevolutionOpticalManipulation2003,gordonFutureProspectsBiomolecular2022} to measure small forces at piconewton to femtonewton regimes \cite{xinOpticalForcesFundamental2020}.
Importantly, optical tweezers and  their thermal \cite{tiwariSingleMoleculeSurface2021, sharmaOptothermalPullingTrapping2021,paulOptothermalEvolutionActive2022a,patraPlasmofluidicSinglemoleculeSurfaceenhanced2014, ghoshAllOpticalDynamic2019, ghoshNextGenerationOpticalNanotweezers2020,sharmaLargescaleOptothermalAssembly2020, nalupurackalHydrothermophoreticTrapMicroparticles2022, kumarPitchrotationalManipulationSingle2020, kumarTrappedOutofEquilibriumStationary2020, ghoshDirectedSelfAssemblyDriven2021, kotsifakiRoleTemperatureinducedEffects2022,kotsifakiPlasmonicOpticalTweezers2019, linOptothermoelectricNanotweezers2018} and nonlinear optical variants \cite{ishiharaOpticalManipulationNanoscale2021,deviGeneralizedDescriptionNonlinear2020,deStableOpticalTrapping2009} are routinely used to trap a variety of matter - from individual atoms to micro-organisms \cite{zemanekPerspectiveLightinducedTransport2019}.
They find applications in both fundamental sciences and emerging applications \cite{volpeRoadmapOpticalTweezers2022, jonesOpticalTweezersPrinciples2015}.\

\subsection{Optical Spanners}
In late 1980s, Coullet and coworkers \cite{coulletOpticalVortices1989} were interested in laser cavities that could emit radiation in the form of an optical vortex.
These explorations were mainly in theoretical domain but the emergence of an optical vortices, in analogy to superconducting vortices, were considered and discussed.
An important breakthrough was published in 1992 by Allen and coworkers \cite{allenOrbitalAngularMomentum1992}, who reported on experimental generation of laser beams carrying orbital angular momentum.
The authors also proposed an experiment to measure the torque due to the orbital angular momentum.
This paper laid a foundation for further experimental exploration of optical vortex beams in a variety of geometries and configurations \cite{padgettOrbitalAngularMomentum2017a, shenOpticalVortices302019a, kumarMakingOpticalVortex2011, viswanathanGenerationOpticalVector2009,  kumarInformationContentOptical2011, kapoorOpticalVortexArray2016, xavierSculptured3DTwister2011, bekshaevParaxialLightBeams2008, khareOrbitalAngularMomentum2020, remeshDiffractionOffaxisVectorbeam2022, singhTransverseSpinScattering2018, royManipulatingTransverseSpin2022, kumarTopologicalStructuresVectorvortex2014}.
Around 1994, Babiker et al.,\cite{babikerLightinducedTorqueMoving1994} theoretically studied interaction of optical orbital angular momentum and moving atoms, which hinted on some intriguing effects that may be further explored.
In 1995, He et al., \cite{heDirectObservationTransfer1995} experimentally trapped light-absorbing particles with vortex laser beams.
The trapping location within the beam was at the central intensity minimum of the beam. 
They also observed spinning of these particles due to transfer of angular momentum. Around 1996, Gahagan and Swartzlander \cite{gahaganOpticalVortexTrapping1996} demonstrated optical vortex trapping in three dimensions.
In this study particle of size 20 micrometers were trapped at the central region of the vortex beam.\
In 1996, the concept of optical spanner \cite{simpsonOpticalTweezersOptical1996} was proposed numerically and achievable angular velocities and accelerations were explored.
Later, Simpson and coworkers \cite{simpsonMechanicalEquivalenceSpin1997} experimentally showed how an optical vortex trap can be used as optical spanner.
This further motivated researchers to trap and tweeze low index objects such as water droplets in an organic solvents \cite{gahaganTrappingLowindexMicroparticles1998, gahaganSimultaneousTrappingLowindex1999}.The dependence of experimental parameters such as size of the vortex core and numerical aperture of the objective lens were systematically studied \cite{gahaganTrappingLowindexMicroparticles1998}.
The concept of optically driven micromachine was envisaged by Friese et al. \cite{frieseOpticalAlignmentSpinning1998}, which showed rotation of birefringent particle due to polarization interaction at frequencies above 350 Hz. One of the early studies on metallic particles interacting with optical vortex was done by O'Neil and Padgett \cite{oneilThreedimensionalOpticalConfinement2000}, who observed three-dimensional confinement of metallic colloids in an optical vortex trap.The angular momentum transfer process in terms of spin and orbital components were discussed.\
All these studies have laid an excellent experimental foundation \cite{padgettTweezersTwist2011, padgettOrbitalAngularMomentum2017a, shenOpticalVortices302019a, yaoOrbitalAngularMomentum2011, rubinsztein-dunlopRoadmapStructuredLight2016, volpeRoadmapOpticalTweezers2022, stuhlmullerColloidalTransportTwisted2022, tsujiEffectHydrodynamicInterparticle2020, nakajimaVisualizationOpticalVortex2022,sharmaOpticalOrbitalAngular2019,paulSimultaneousDetectionSpin2022,shiOptomechanicsBasedAngular2016}  for optical vortex to be used as a "optical spanner", which has not only been used to rotate trapped object on its axis, but also revolve in an orbit with controlled angular velocities\cite{otteOpticalTrappingGets2020,yangOpticalTrappingStructured2021}.\
\section{Optical forces and torques on plasmonic colloids}
Conservation of linear and angular momentum of light is central to computation of optical forces on matter.
In the context of optical forces exerted by a light beam on matter, one can classify the interaction based on size parameter  $r = 2\pi a/\lambda$ (for a detailed description on this, see chapter 1 of the book by Jones et al \cite{jonesOpticalTweezersPrinciples2015}).
Here $a$ is the characteristic dimension of the object interacting with a light beam of wavelength $\lambda$.
If $r<<1$, then one can use dipole approximation to treat the interaction problem, and compute the forces accordingly.
Forces on atoms, small molecules and certain nanoparticles can be studied in this regime.
If $r>>1$, then the geometrical optical treatment can be used to compute the forces. 
This is applicable to micron sized colloids and biological cells. 
In case of $r\simeq1$, a full electromagnetic treatment based on Maxwell stress tensor is necessary to compute the optical forces.
Depending upon the size of plasmonic nano-colloids \cite{svobodaOpticalTrappingMetallic1994, lehmuskeroLaserTrappingColloidal2015, maragoOpticalTrappingManipulation2013, dienerowitzOpticalManipulationNanoparticles2008, spesyvtsevaTrappingMaterialWorld2016}, dipole approximation method or Maxwell stress-tensor method is used to compute optical forces.

\subsection{Dipole Approximation}
 
In this case, a nano-colloid is assumed to be a point dipole.
When illuminated, the electric field of light induces a dipole moment in the colloid.
This induced dipole $\boldsymbol{p}$ is related to electric field $\boldsymbol{E}$ via polarizability tensor $\boldsymbol{\alpha}$, and is given by the expression
\begin{equation}
\boldsymbol{p}=n_m^2 \boldsymbol{\alpha} \boldsymbol{E} ,
\end{equation}
where $n_m$ is the index of the medium in which colloid is immersed.

The Lorentz force sets the colloid into motion.
Specifically, the force can be divided in to three components.
First is the gradient force, $F_{\mathrm{grad}}(r)$. The second is the extinction force, $F_{\mathrm{ext}}(r)$. The third is a scattering force due to curl of the spin angular momentum of light, whose implication are under exploration nowadays \cite{xinOpticalForcesFundamental2020,nieto-vesperinasOpticalTorqueElectromagnetic2015}.
The first two forces have been extensively studied, and are given by the expression \cite{jonesOpticalTweezersPrinciples2015}: 

\begin{equation}\label{fgrad}
F_{\mathrm{grad}}(r)=\frac{1}{2 \varepsilon_0} \operatorname{Re}\{\alpha\} \nabla I(\mathbf{r})
\end{equation}

\begin{equation}\label{fext}
F_{\mathrm{ext}}(\boldsymbol{r})=\frac{n_{\mathrm{m}}}{c} \sigma_{\mathrm{ext}}\langle S(r)\rangle
\end{equation}

where $\nabla I(\mathbf{r})$ is the gradient of intensity, $\sigma_{\mathrm{ext}}$ is the extinction cross section and $\langle S(r)\rangle$ is the time averaged Poynting vector.
The extinction scattering is the sum of absorption cross-section, $\sigma_{\mathrm{abs}}$, and scattering cross section $\sigma_{\mathrm{scat}}$:

\begin{equation}
\sigma_{\mathrm{ext}} = \sigma_{\mathrm{abs}} + \sigma_{\mathrm{scat}}.
\end{equation}

The cross-sections themselves are dependent on parameters including frequency of light, $\omega$, and real and imaginary parts of ${\alpha}$ , and are given by equations :

\begin{equation}
\sigma_{\mathrm{abs}}=\frac{\omega n_{\mathrm{m}}}{c \varepsilon_0} \operatorname{lm}\{\alpha\},
\end{equation}

and 
\begin{equation}
\sigma_{\mathrm{scat}}=\frac{\omega^4 n_{\mathrm{m}}^4}{6 \pi c^4 \varepsilon_0^2}|\alpha|^2 .
\end{equation}

The gradient force (equation \ref{fgrad}) is generally attractive, and pulls the colloids towards the intensity maxima of a laser beam.
This is the basis of a single beam optical tweezer \cite{ashkinObservationSinglebeamGradient1986}, and is facilitated by tightly focused laser beam using a high numerical aperture objective lens.
The extinction force (equation \ref{fext})  has contribution due to scattering and absorption and is generally along $\langle S(r)\rangle$, that is the propagation direction of the beam.
In addition to these two forces there is a spin-curl force \cite{albaladejoScatteringForcesCurl2009}, $F_{\mathrm{sc}}(r)$, and is given by the expression \cite{jonesOpticalTweezersPrinciples2015} :

\begin{equation}
F_{\mathrm{sc}}(\boldsymbol{r})= -\frac{1}{2}  c  \sigma_{\mathrm{ext}} (\nabla\times s),
\end{equation}

where $s$ is the time averaged spin density, and $c$ is the speed of light in vacuum. The spin-curl force is non-conservative scattering contribution, and is mainly due to gradients of polarization. On a relative scale, their contribution towards trapping is negligible, but there is more to learn on this component.

\subsection{ Maxwell stress tensor} \label{mst}

If the particle size is of the order of wavelength of light, then dipole approximation cannot be used.
In such a situation, the electromagnetic interaction takes the center stage, and the forces are computed using electromagnetic fields.
The Maxwell's stress-tensor $\overline{\mathrm{T}}_{\mathrm{M}}$ is used for this purpose, and the expression for optical force, $\mathbf{F}_{\mathrm{opt}}$, is given by \cite{jonesOpticalTweezersPrinciples2015} :

\begin{equation}
\mathbf{F}_{\mathrm{opt}}=r^2 \oint_{\Omega} \overline{\mathrm{T}}_{\mathrm{M}} \cdot \hat{\mathbf{r}} d \Omega,
\end{equation}

where the surface integral is taken over a spherical surface of radius $r$ and solid angle, $\Omega$, with the object of interest is at the center. The $\hat{\mathbf{r}}$ is the radial unit vector. 
The $\overline{\mathrm{T}}_{\mathrm{M}}$ depends on the total and scattered electromagnetic fields.
So the main task is to compute these fields for a given geometry, and then utilize them to calculate forces or torques.
 A good discussion on Maxwell stress-tensor and optical forces can be found in the book of Jones et. al (see chapter 5) \cite{jonesOpticalTweezersPrinciples2015}.

\subsection{Optical torques}

Light can carry spin and orbital angular momentum \cite{yaoOrbitalAngularMomentum2011, sipova-jungovaNanoscaleInorganicMotors2020, bruceInitiatingRevolutionsOptical2021}.
This leads to rotation of an object either on its own axis, or around a point in an orbit \cite{lehmuskeroPlasmonicParticlesSet2014, sipova-jungovaNanoscaleInorganicMotors2020}.
Torque due to spin is generally termed as 'intrinsic' torque, and is mainly due to scattering or absorption of light \cite{frieseOpticalAngularmomentumTransfer1996, frieseOpticalAlignmentSpinning1998}.
If an object is aspherical, then geometrical anisotropy leads to rotation \cite{hajizadehBrownianFluctuationsOptically2017, liawRotatingAuNanorod2014}.
A torque is also generated  due to structure of light, as in helical phase of an optical vortex, which sets the object into rotation \cite{oneilIntrinsicExtrinsicNature2002}.
In the electromagnetic regime, the optical torque, $\mathbf{T}_{\mathrm{opt}}$, can be computed using Maxwell stress tensor (see section \ref{mst}), and is given by the expression \cite{jonesOpticalTweezersPrinciples2015} :

\begin{equation}
\mathbf{T}_{\mathrm{opt}}=-r^3 \oint_{\Omega}\left(\overline{\mathrm{T}}_{\mathrm{M}} \times \hat{\mathbf{r}}\right) \cdot \hat{\mathbf{r}} d \Omega.
\end{equation}

The implication of intrinsic and extrinsic torque is a topic of current research \cite{bruceInitiatingRevolutionsOptical2021}. A detailed theoretical framework from conservation of angular momentum has been laid out recently \cite{nieto-vesperinasOpticalTorqueElectromagnetic2015}. 
New prospects such as negative optical torque \cite{dogariuOpticallyInducedNegative2013} have added further impetus to this exploration.

\section{Optical Vortex}
Apart from linear momentum, a beam of light can carry angular momentum.
This can be in the form of spin or orbital angular momentum.
Optical beams with helical phase-fronts have azimuthal component of Poynting vector, which facilitates the orbital angular momentum.
This orbital component circulates with respect to the beam axis, and hence such an optical beam is called as an "optical vortex". 
The central part of an optical vortex has a phase singularity, which results in zero intensity at the center. 
Therefore, such an optical beam has zero amplitude at center with a circulating phase around it. 
Experimentally, the first concrete realization of an optical vortex beam carrying orbital angular momentum was achieved in 1992\cite{allenOrbitalAngularMomentum1992}. 
Since then, optical vortex beams\cite{padgettLightOrbitalAngular2004, allenOpticalAngularMomentum2003, padgettLightTwistIts2000}  and their applications\cite{padgettOrbitalAngularMomentum2017a, shenOpticalVortices302019a} have emerged as one of the most active areas in optics and photonics\cite{senthilkumaranSingularitiesPhysicsEngineering2018, khareOrbitalAngularMomentum2020, andrewsAngularMomentumLight2012, andrewsStructuredLightIts2011}.

\subsection{Gaussian and Laguerre-Gaussian Beams}

\begin{figure}[h]
\centering
\includegraphics[width=12cm]{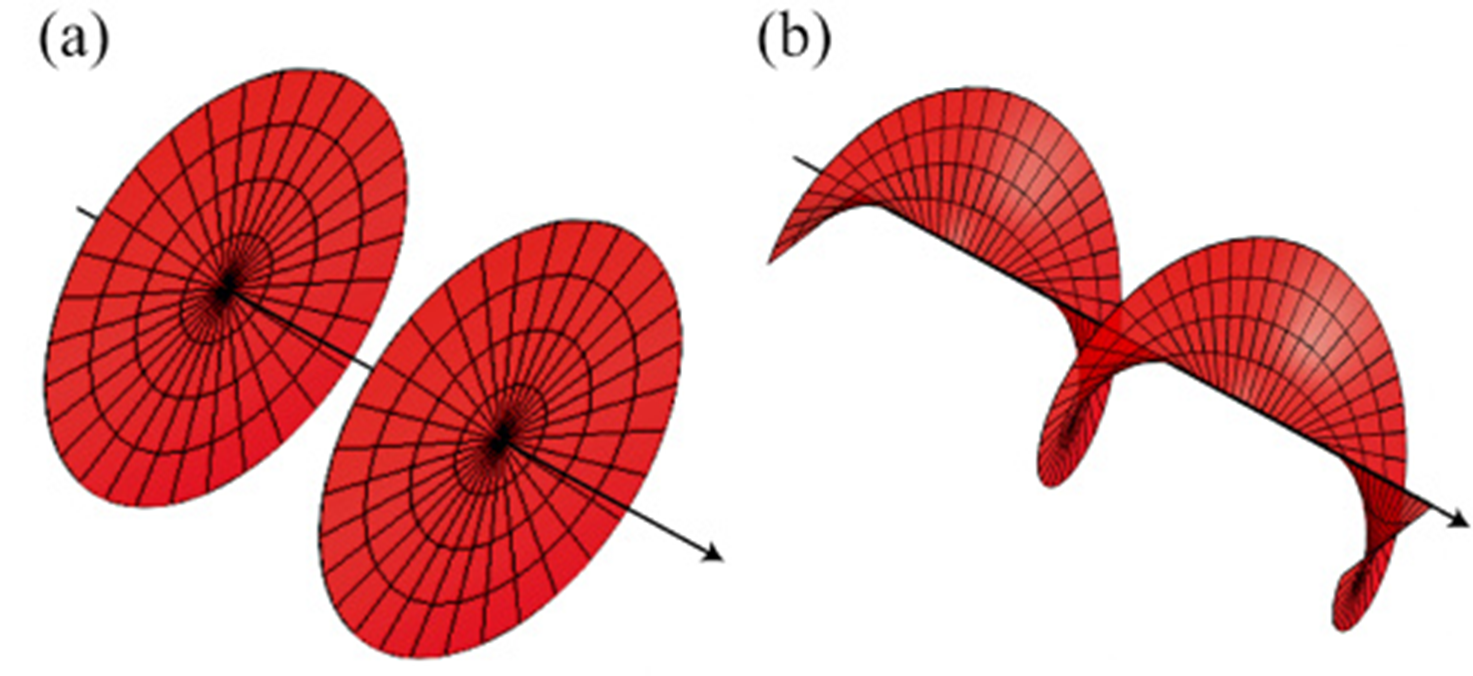}
\caption{Wavefront of (a) a Gaussian beam and (b) an optical vortex beam. Figure reproduced from reference \cite{padgettOrbitalAngularMomentum2017a}}
\label{gnov}
\end{figure}

Under paraxial approximation\cite{salehFundamentalsPhotonics2012}, one of the solutions to Helmholtz wave equation is a Gaussian beam and a more generalized solution is the Laguerre-Gaussian (LG) beam (see chapter 3 of \cite{salehFundamentalsPhotonics2012} for an lucid discussion on optical beams). Parameters of optical vortices are defined by LG beams. Below are the expressions for Gaussian and LG beams in cylindrical coordinates \cite{salehFundamentalsPhotonics2012}. An important difference between the two beams is the phase factor, which has implications in defining the helical phase front of an LG beam (see figure \ref{gnov}).

The amplitude of a Gaussian beam is given by  :
\begin{equation}\label{G beam}
U(\mathbf{r})=A_0 \frac{W_0}{W(z)} \exp \left[-\frac{\rho^2}{W^2(z)}\right] \exp \left[-i k z-i k \frac{\rho^2}{2 R(z)}+i \zeta(z)\right]
\end{equation}
where $W(z)$ is the beam width, $R(z)$ is wavefront radius of curvature, $q(z)$ is called the q-parameter of the beam, $z_0$ is known as the Rayleigh length. This is the distance from the beam waist along the propagation direction at which the cross section of the beam is doubled, and $A_0=A_1 / i z_0$ and $A_1$ is a constant. They are related to each other by the following equations :

\begin{equation}
\begin{aligned}
W(z) &=W_0 \sqrt{1+\left(\frac{z}{z_0}\right)^2} \\
R(z) &=z\left[1+\left(\frac{z_0}{z}\right)^2\right] \\
\zeta(z) &=\tan ^{-1} \frac{z}{z_0} \\
W_0 &=\sqrt{\frac{\lambda z_0}{\pi}}
\end{aligned}
\end{equation}

A more general set of solutions to paraxial Helmholtz equation is given by the LG beam, whose amplitude is given by

\begin{equation}\label{LG beam}
\begin{aligned}
U_{l, m}(\rho, \phi, z) &=A_{l, m}\left[\frac{W_0}{W(z)}\right]\left(\frac{\rho}{W(z)}\right)^l \mathbb{L}_m^l\left(\frac{2 \rho^2}{W^2(z)}\right) \exp \left(-\frac{\rho^2}{W^2(z)}\right) \\
& \times \exp \left[-i k z-i k \frac{\rho^2}{2 R(z)}-i l \phi+i(l+2 m+1) \zeta(z)\right]
\end{aligned}
\end{equation}

where $\mathbb{L}_m^l(\cdot)$ is the generalized Laguerre polynomial function \cite{gburSingularOptics2016, gburMathematicalMethodsOptical2011}. The topological charge of the the LG beam is  $\ell$.
It is the azimuthal index carrying an orbital angular momentum of $\ell \hbar$ per photon. The  $p$ represents the number of nodes in the intensity distribution. Notice that if  $\ell = 0$ in equation \ref{LG beam}, then we recover the Gaussian beam given by equation \ref{G beam}.\
In the context of rotation of a Brownian object due to an optical vortex beam, it is the value of  $\ell$ which determines the speed of orbital rotation, as will be discussed later.

\section{ Plasmonic colloids interacting with optical vortex }
\subsection{ Orbital rotation of gold nano-colloids}

\begin{figure}[h]
\centering
\includegraphics[width=12cm]{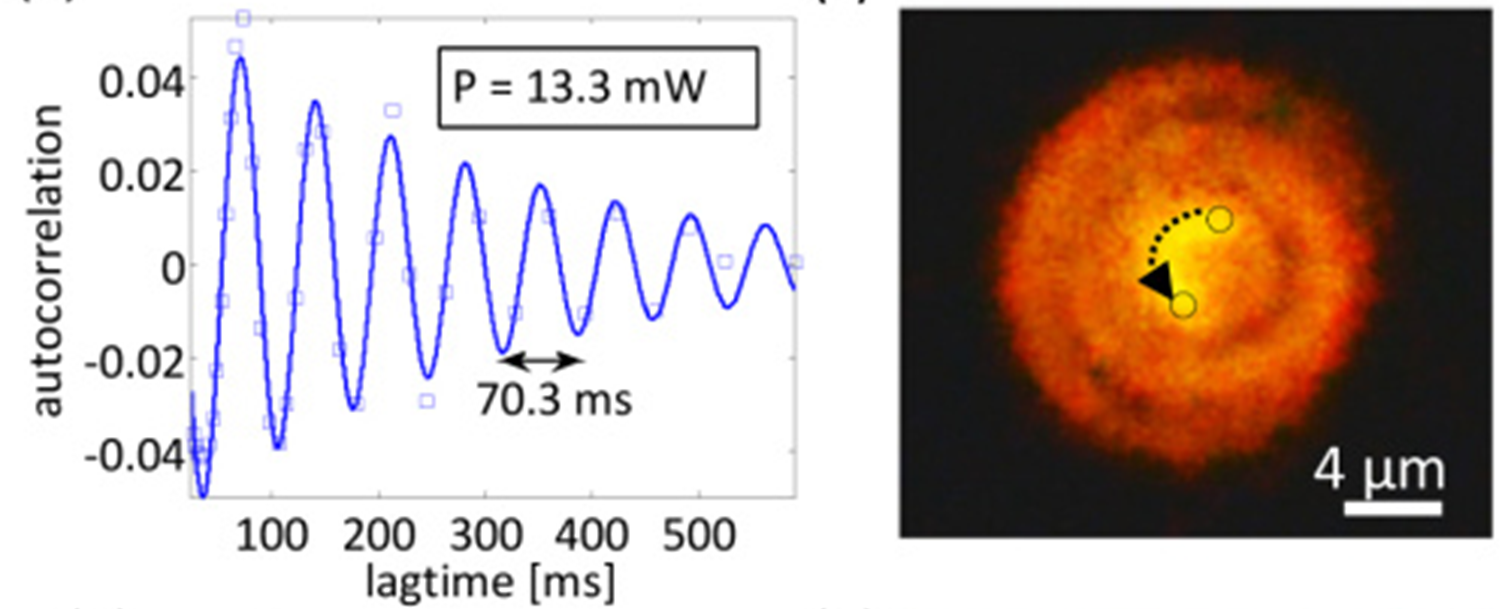}
\caption{The left panel shows autocorrelation of rotation of gold nano-colloids in an optical vortex trap. The right panel is a dark-field image of the changing position of a nano-colloid. Figure reproduced from reference \cite{lehmuskeroPlasmonicParticlesSet2014}}
\label{fast}
\end{figure}

Tightly focused optical vortices can trap and rotate plasmonic nano-colloids at high speeds.
The phase component of the vortex trap, including the sign of the topological charge $l$, can be harnessed to control the handedness of rotation.
Lehmuskero et al., \cite{lehmuskeroPlasmonicParticlesSet2014} were able to utilize an infrared optical vortex ($l = 8$) trap at 830 nm to induce orbital motion on individual gold nano-colloids.
They observed velocity upto 1.1 millimeter per second, which is significant and useful for various applications including microfluidics and micro-rheology. 
Given that gold nano-colloids are highly polarizable, a large intensity gradient force can be applied, and hence they can be attracted towards maximum intensity of the optical vortex.
The scattering and absorption from the nano-colloids too have a significant role in this interaction.
These optical properties collectively contribute to the radiative torque. 
Significant absorption can also lead to heating of the colloids, which in turn increases the temperature of the surrounding fluid medium.
This further influences the viscosity and plays a role in imparting a Stokes drag, and thus a viscous torque.
For the orbital motion to sustain, the radiative torque and viscous torque have to balance each other out.
Depending upon the composition of the colloid, properties of the medium and parameters of the vortex, the resultant torque and angular velocity can be influenced.
In order to quantify the orbital motion, the authors \cite{lehmuskeroPlasmonicParticlesSet2014} performed auto-correlation analysis of the scattered light from the circulating nano-colloid (see figure \ref{fast}, right panel).
Interestingly, they were able to capture the smeared trajectory of the circulating colloid (see figure \ref{fast}, left panel) in the orbit. 
This study shows how individual plasmonic colloids can be set into fast orbital motion using vortex beams, and can have further implications in utilizing orbital angular momentum to set nano-object into controlled rotational motion.

\subsection{ Spin-orbit coupling in rotation}

\begin{figure}[h]
\centering
\includegraphics[width=12cm]{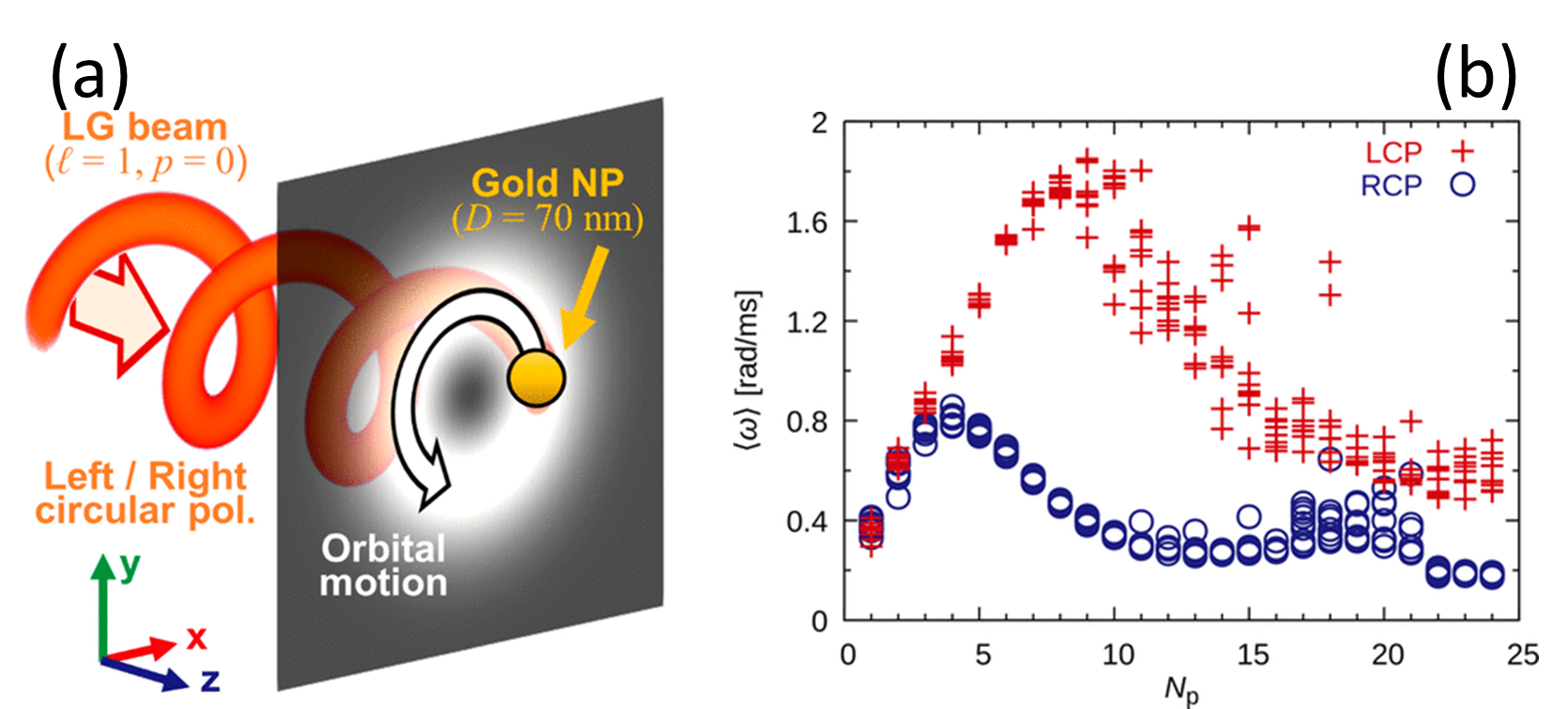}
\caption{(a)Schematic representation of circularly polarized optical vortex interacting with a gold colloid. (b) Comparison of angular velocity vs number of colloids for two opposite handedness of the circularly polarized vortex beams. Figure reproduced from reference \cite{tamuraInterparticleInteractionMediatedAnomalousAcceleration2019}}
\label{so}
\end{figure}

Coupling between the spin and orbital degree of freedom has been of interest for a long time \cite{padgettOrbitalAngularMomentum2017a}.
Optical beams carrying these momenta can give rise to interesting spin-orbit effects \cite{bliokhGeometricalOpticsBeams2006, bliokhGeometrodynamicsSpinningLight2008, bliokhSpintoorbitalAngularMomentum2011, bliokhSpinOrbitInteractions2015}.
This has opened up new avenues in photonics \cite{bliokhSpinOrbitInteractions2015,  eismannTransverseSpinningUnpolarized2021}, including quantum optical regimes \cite{bliokhQuantumSpinHall2015}.
The interaction of spin and orbital angular momentum in an optical vortex can also be used to control the dynamics of plasmonic colloids.
Recently Tamura et al. \cite{tamuraInterparticleInteractionMediatedAnomalousAcceleration2019}, have theoretically shown how such coupled interactions can be utilized to control the motion of gold nano-colloids (see figure \ref{so}(a)).
Specifically, they show how light induced forces between gold colloids (70nm diameter) can play a critical role in influencing the orbital dynamics when driven by optical vortex beam at 1064nm wavelength.
By using Brownian dynamics based simulations, they show acceleration and deceleration of orbital angular velocity depending upon how the spin and orbital parts of the angular momentum are coupled to each other.
For $l = 1$ beam, they observe acceleration with left circular polarization, and a deceleration with right circular polarization (see figure \ref{so}(b)).
Furthermore, they show how inter-colloid interaction influences the arrangement of colloids in the vortex, and this interaction depends on the spin-orbit coupling.
Such intriguing spin-dependent dynamics can indeed be of great use when one needs to spatially control the rotational speeds and arrangement of colloids.        

\subsection{  Converting spin to orbital motion via multiple scattering }

\begin{figure}[h]
\centering
\includegraphics[width=6cm]{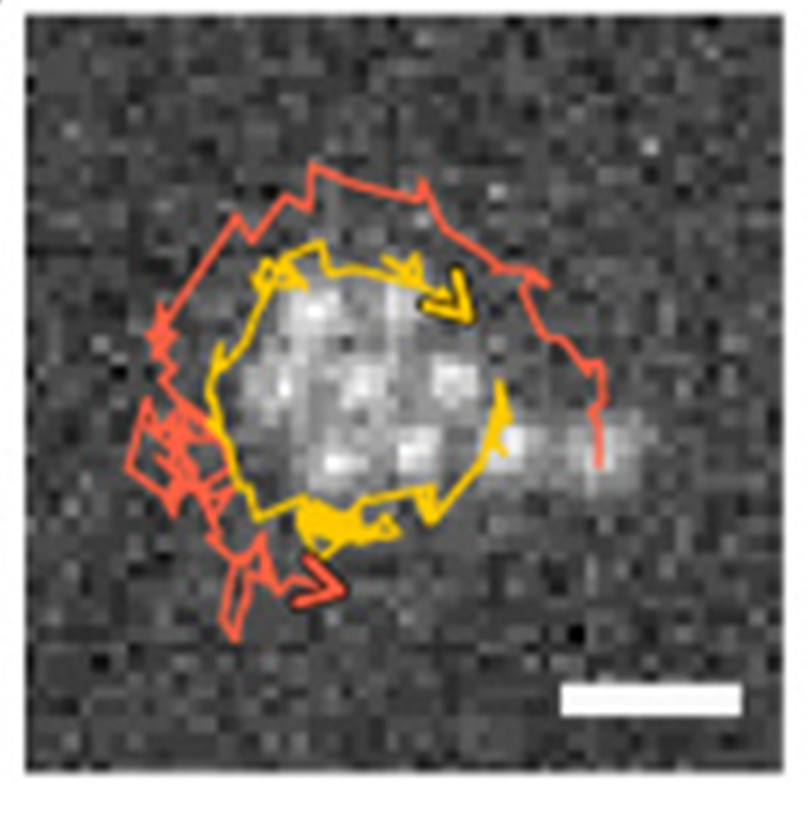}
\caption{Silver nano-colloidal probe ($100$ nm) orbiting a spinning hexagonal array of silver colloids realized by circularly polarized optical trap.  The handedness of the orbital motion of the outer colloid is opposite to the rotation of the array, as indicated by the coloured trajectories and arrows. Scale bar is 1 $\mu m$. Figure reproduced from reference \cite{parkerOpticalMatterMachines2020}}
\label{gear}
\end{figure}

When multiple plasmonic nano-colloids are optically assembled at close proximity (within a distance of excitation wavelength), the light scattered from the colloidal system can show intriguing optical effects.
An array of plasmonic colloids, under appropriate conditions, can facilitate non-equilibrium effects such as negative optical torques.
Parker et al. \cite{parkerOpticalMatterMachines2020}, used the multiple scattering effects of trapped plasmonic colloids to experimentally realize 'optical matter machine'.
They trap and drive a hexagonal array of silver nano-colloids ($7 5$ nm radius) with a 'circularly polarized optical trap' using a $800$ nm laser beam (see figure \ref{gear}).
As a consequence of the spin angular momentum of the rotating array, an orbital angular momentum is generated which can lead to orbital motion of a nanoscale probe object, such as a 100 nm silver nano-colloid (see outer nano-colloid in figure \ref{gear}).
In this study, the collective scattering modes of the array plays a critical role of providing an optical torque to the orbiting object, and specifically this orbital torque has a handedness that is opposite to that of the circularly polarized beam.
The authors use the analogy of a planetary gear system, and view the realized system from a framework of optical matter machines.
This work collectively shows how spin and orbital angular momentum can co-exist in a system, and how plasmonic colloidal arrays with their enhanced scattering can play a vital role.  

\subsection{ Rotation of a plasmonic colloidal nanowire}

\begin{figure}[h]
\centering
\includegraphics[width=12cm]{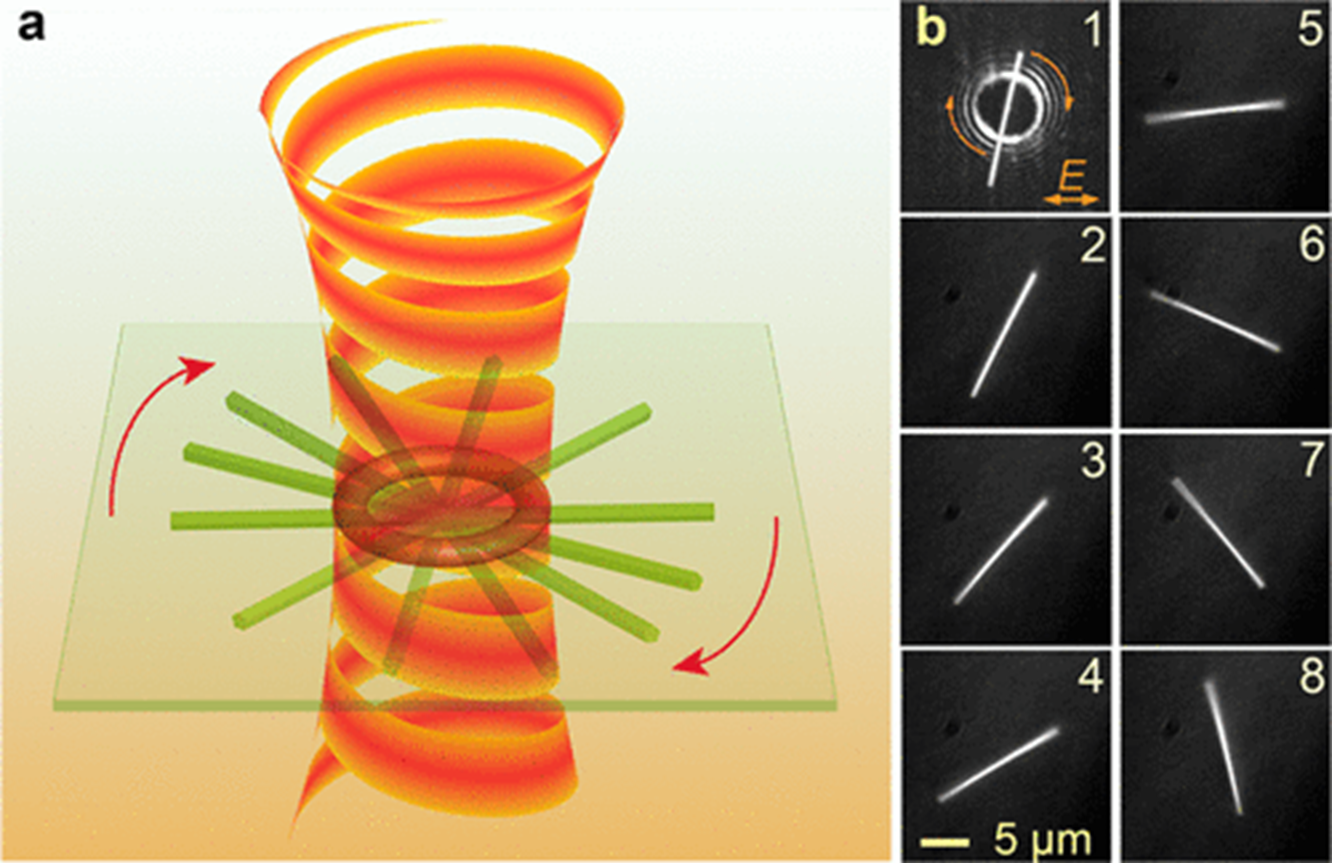}
\caption{(a)Schematic of nanowire rotation under the influence of an optical vortex beam. (b)Time-series of dark-field optical images of a rotating silver nanowire driven by an optical vortex beam. The top image shows the position of unfiltered vortex beam with respect to the nanowire.  Figure reproduced from reference \cite{yanOpticalVortexInduced2013}}
\label{AgNW}
\end{figure}

The transfer of angular momentum to rotate geometrically anisotropic objects in a fluid has applications in soft-matter science and micro-fluidic technologies.
An interesting prospect is to utilize optical vortex beam to manipulate elongated objects such as nanowires.
Specifically, nanowires made of metals such as silver and gold can facilitate surface plasmons, and hence can strongly interact with visible light.
In 2013, Yan et al., \cite{yanOpticalVortexInduced2013} utilized optical vortex beam to manipulate a silver colloidal nanowire (see figure \ref{AgNW} a).
They clearly showed how the nanowire can be rotated (see figure \ref{AgNW} b) with respect to an imaginary axis by inducing torque on different parts of the nanowire. 
The topological charge of the vortex beam in this case was $l = 20$ and this index could be used to control the rotation of the nanowire.
It was observed that as the value of $l$ increased, the rotation frequency increased, thus providing a controllable parameter to impart torque. 
Furthermore, the sign of the topological charge could be changed to flip the handedness of the rotation, which added another parameter to control rotation.
The authors \cite{yanOpticalVortexInduced2013} also examined the nature of torque which was imparted on the nanowire.
Apart from the contribution due to orbital angular momentum and fluid drag (which competes with the optical torque), the plasmonic contribution was emphasized. 
This work motivates further questions to explore the role of plasmon resonance in effective transfer of orbital angular momentum from a vortex beam to a colloidal object.
Given that a variety of anisotropic plasmonic colloids \cite{sipova-jungovaNanoscaleInorganicMotors2020} can be prepared using chemical synthesis, the interaction between optical vortex and colloids can be systematically studied. 
There is a lot more to learn from such studies.

\subsection{ Resonant confinement of plasmonic colloids }

\begin{figure}[h]
\centering
\includegraphics[width=14cm]{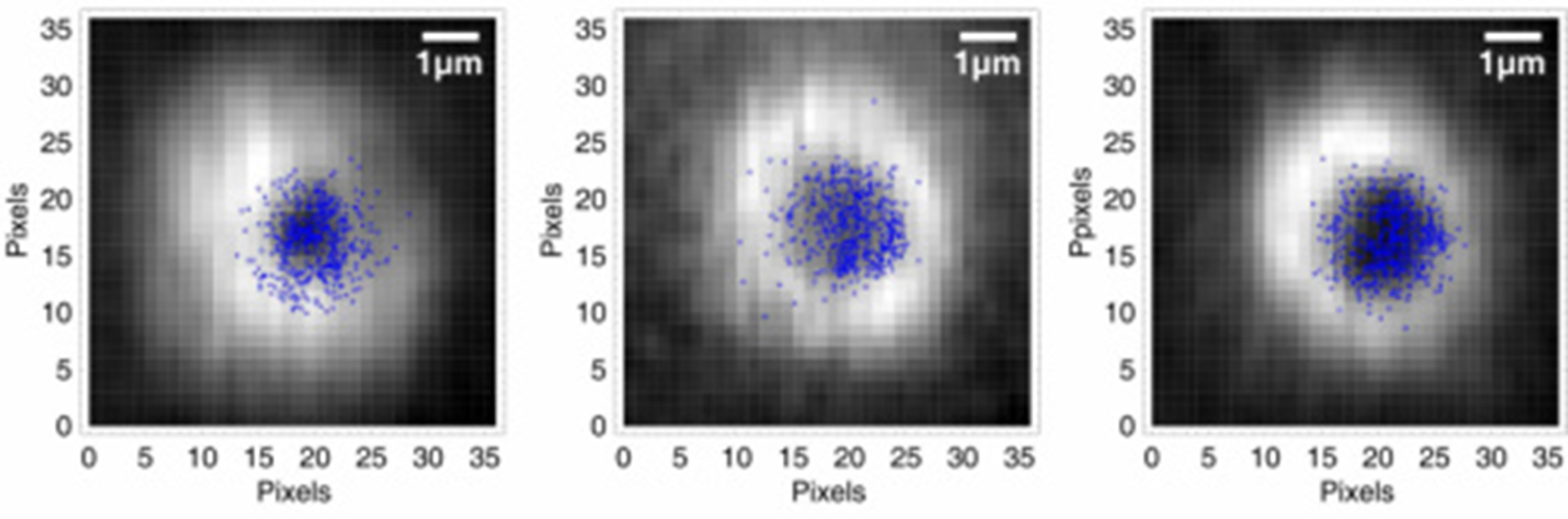}
\caption{Spatial data points tracing the movement of a single gold nanoparticle confined to a vortex beam and various wavelengths a) 488nm, b) 514nm  and c) 528nm. Figure reproduced from reference \cite{dienerowitzOpticalVortexTrap2008}}
\label{confine}
\end{figure}

As discussed earlier, and optical vortex beam has zero intensity at its center.
This means a nanoparticle which is interacting with the beam is subject to a differential forces at various points with in an optical beam.
This differential force field, unlike a Gaussian beam, is not continuous in intensity due to the singularity at the center of an optical vortex.
As a result, a resonant nano-colloid can be potentially confined to the region of minimum intensity within an optical vortex.
This is exactly what Dienerowitz et al., \cite{dienerowitzOpticalVortexTrap2008} showed in their experiment (see figure \ref{confine}).
They employed resonant optical vortex beams to interact with plasmonic gold or silver nano-colloids.
The wavelengths of the trap were chosen either at 528nm or  514nm or 488nm.
These values overlap with different parts of the plasmonic resonances of the gold colloids.
As a result of the resonance overlap, three components of forces come in to play \cite{dienerowitzOpticalVortexTrap2008} : first is an attractive gradient force, the second is a repulsive scattering force, and third is a repulsive force that depends on the polarizability of the metallic colloids. 
The dominant contribution is due to the scattering force, which pushes the colloids towards the region of minimum intensity and hence confines them towards the center of an optical vortex.
The authors also found that the efficiency of confinement of plasmonic colloids depend on size.
Smaller the nano-colloid (around 70nm) better is the confinement.
Thus, depending upon the material property and size of the plasmonic colloid, the forces can vary, and hence the confinement can be controlled.
In addition to the confinement effect, the authors \cite{dienerowitzOpticalVortexTrap2008} were able to transfer the orbital angular momentum of the optical vortex to the plasmonic nano-colloids.
They observed rotation of the trapped colloids at the periphery of the intensity minimum of the vortex.
This study highlights the capability of a resonant vortex beam to create complex potentials to manipulate plasmonic colloids.

\subsection{ Diffusion dynamics : vortex and their lattices}

\begin{figure}[h]
\centering
\includegraphics[width=12cm]{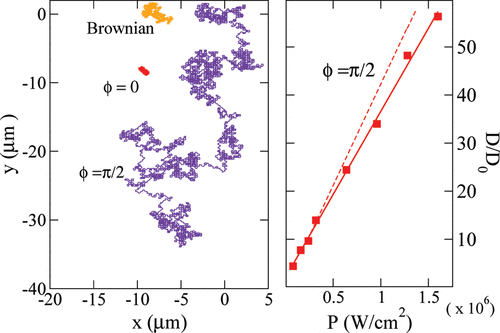}
\caption{Diffusion of gold nano-colloid in optical vortex lattice. The left panel compares the trajectory between conventional Brownian motion, standing wave lattice formed by interfering light with phase difference of $\phi = 0$ and $\phi =  90$ degrees. In the right panel, the deviation of diffusivity from Brownian dynamics is shown (dotted line is for conventional Brownian motion). Figure reproduced from reference \cite{albaladejoGiantEnhancedDiffusion2009}}
\label{giant}
\end{figure}

Diffusion dynamics of colloids is one of the most important concept in soft-matter science and technology \cite{bundeDiffusiveSpreadingNature2017}.
By studying how colloidal diffusion is influenced by local environment gives an insight into equilibrium and non-equilibrium conditions of the surrounding including micro-mechanical properties \cite{grierOpticalTweezersColloid1997}.
Transport properties of colloids can also be influenced by applying an external field, of which optical potentials provide a versatile route. 
Conventionally, conservative optical forces \cite{grierOpticalTweezersColloid1997, bradshawManipulatingParticlesLight2017} are utilized to influence the colloidal dynamics and transport. 
In recent times, non-conservative optical forces \cite{sukhovNonconservativeOpticalForces2017, roichmanInfluenceNonconservativeOptical2008, wuDirectMeasurementNonconservative2009} have been explored in the context of driving colloids out of equilibrium. 
In the context of optical vortex, the helical phase front of the beams can facilitate non-conservative optical forces \cite{sukhovNonconservativeOpticalForces2017}.
\\ An interesting prospect is to ask : how can such a non-conservative force be harnessed to influence dynamics of plasmonic colloids? 
In 2009, Albaladejo et al., \cite{albaladejoGiantEnhancedDiffusion2009} numerically explored the problem of gold nano-colloids interacting with a lattice of optical vortices (consult reference \cite{zhuOpticalVortexLattice2021} for a recent review on optical vortex lattice).
They created a standing wave of optical vortices by combining cross-polarized counter-propagating waves, and studied the diffusion dynamics of gold colloids ($50 nm$ size range).
As a result of this theoretical set-up, the authors found enhanced diffusion of gold colloids under the influence of this non-conservative force field when compared to force fields that of a conventional optical tweezers. 
\\As a follow up to this study, silver nano-colloids were explored \cite{albaladejoLightControlSilver2011}, and contrasted to gold nano-colloids.
Depending on the  excitation of plasmon resonance wavelength, three different regimes of dynamics were identified.
The first one was a trapped configuration, where closed trajectories were observed.
In the second configuration, as the excitation wavelength was matched towards plasmonic resonance of the silver nano-colloids, a drastic increment in the diffusion constant was observed.
In the third regime, at the limit of longer wavelength compared to plasmon resonance, the authors simulated quasi-one dimensional trap configuration.
This specific one-dimensional configuration may have implication in Brownian ratcheting of colloids leading to directional transport, and needs further experimental exploration.\\  
In a separate study, Zapata et al. \cite{zapataControlDiffusionNanoparticles2016} explored the general framework in which diffusion of colloids can be controlled with a lattice of optical vortex.
They identified flow patterns for the force field that includes conservative dipolar component and the non-conservative contribution from the optical vortex.
For the computed flow paths, they determined probable escape pathways for the colloids in the vortex lattice.
By further tuning the force fields, the authors envisage that the colloidal motion can be deterministically varied, which can be further used in applications such as nano-fluidics.\\
Emergence is a fascinating concept and this can apply to assembly and dynamics of colloids under optical perturbation\cite{dobnikarEmergentColloidalDynamics2013}.
Specifically, optical vortex and their lattices can lead to collective dynamics of colloids that are unconventional.
Such collective dynamics of gold nano-colloids in vortex lattices has been explored in the theoretical domain \cite{delgado-buscalioniEmergenceCollectiveDynamics2018}.
The authors observe synchronized motion of colloids and swarm formation when driven by optical vortex fields. 
Unidirectional motion of nano-colloids with speeds reaching $cm/s$ have been numerically calculated.\\
In another study, the prospects of diffusion of gold nano-colloidal dimer under the influence of optical vortex lattice has been explored \cite{melendezOptofluidicControlDispersion2019}, where the breaking of spherical symmetry in the geometry leads to interesting effects in the diffusion.
The authors contrast the dynamics of dimers with monomers, and theoretically quantify their diffusion characteristics. 
Furthermore, the dimer gap length (controlled by polymer chain) is used as a parameter to explore the effect on diffusion properties.\\
To summarize, there is a rich prospect of controlling diffusion dynamics of plasmonic colloids using optical vortex and their lattice. A majority of the studies until now are driven by theoretical ideas and numerical simulations. There is an impetus to explore these problems in experimental domain.

\section{Prospects and Conclusion}
Vortex beams can carry fractional topological charges \cite{taoFractionalOpticalVortex2005}.
The prospects of optical trapping with such fractional vortex beams is an interesting area that needs further exploration.
Conventionally, vortices that are used for trapping experiments are limited by optical diffraction.
Going beyond this limit, nano-vortices can facilitate some interesting opportunities.
David and coworkers\cite{davidTwodimensionalOpticalNanovortices2016} have shown how nano-fabricated silicon waveguides can be used to produce vortices as small as $60$ nm, with desired topological charge.
Such nano-vortices can facilitate interesting optical potentials and need further exploration in the context of nanomanipulation.
In the context of experiments with structured light, Rodrigo, Alieva and coworkers \cite{rodrigoLightdrivenTransportPlasmonic2016, rodrigoProgrammableOpticalTransport2018} have been studying on-demand transport of plasmonic colloids using structured optical beams including optical knots and networks \cite{rodrigoProgrammableOpticalTransport2018}.
They have also been exploring \cite{rodrigoAllopticalMotionControl2021, rodrigoTailoredOpticalPropulsion2020} propulsion mechanism in optothermal field facilitated by structured optical traps including three dimensional optical potentials.
The opportunities of optical vortex with spheroidal plasmonic resonant nanostructures have also been under theoretical exploration \cite{grigorchukLaserinducedAngularMomentum2018} in recent times, including topics such as inducing rotation of such metal nanoparticles under the excitation of ultrashort pulses. The electromagnetic interaction in such situations is strongly influenced by the underlying plasmonic resonances of the metal particle and hence the coupling can be sensitive to the excitation wavelength and polarization.
Other theoretical prospects such as Newtonian orbits of nano-colloids in structured optical fields \cite{ferrer-garciaNewtonianOrbitsNanoparticles2019}, rotation of optically bound assembly by spin-orbit interaction of light \cite{taoRotationOpticallyBound2021} have also gained prominence in recent times.

Thus, we have seen how optical vortices and their lattice can interact and influence the confinement and dynamics of plasmonic colloids. The aim of this review was to highlight some emerging trends in the context of the mentioned interaction. Of course, this is not a comprehensive assessment of the literature, but we hope that it will motivate researchers to study the discussed subject in greater detail. Structured light and its interaction with matter is a fascinating area of research, and optical vortex-colloid interaction is a small subset of this. We envisage that more interesting work will emerge in this direction, both from fundamental and application viewpoint. 

\begin{acknowledgements}
The author thanks his group members (past and present) for various discussion on plasmonic colloids and structured light. He also acknowledges financial support from the Swarnajayanti fellowship grant (DST/SJF/PSA-02/2017–18) from DST-India  and AOARD grant (FA2386-22-1-4017).
\end{acknowledgements}

\bibliography{ijp}

\end{document}